\documentclass[aps,prl,amsmath,twocolumn,groupedaddress,letterpaper,floatfix]{revtex4-2}
\usepackage{graphicx,color,soul}
\usepackage{verbatim}
\usepackage{amssymb}   
\usepackage{amsmath}
\usepackage{amsfonts}
\usepackage{mathdots}
\usepackage{hyperref}
\usepackage{epsfig}
\usepackage{braket}
\usepackage{bm}

\begin{document}
\title{Singlet-triplet mixed state in spin-orbital-parity coupled superconductors}
\author{Tsz Fung Heung}
\thanks{These two authors contributed equally}
\affiliation{Department of Physics, Hong Kong University of Science and Technology, Clear Water Bay, Hong Kong, China} 
\author{Yip Chun Wong}
\thanks{These two authors contributed equally}
\affiliation{Department of Physics, Hong Kong University of Science and Technology, Clear Water Bay, Hong Kong, China} 

\date{\today}
\begin{abstract}
 
Unlike noncentrosymmetric superconductors, the effects of spin singlet- and spin triplet-pairing mixing in centrosymmetric superconductors remain ambiguous. Recently, it was experimentally demonstrated that  an anisotropically enhanced in-plane critical fields beyond Pauli limit would be induced by the coupling of spin,orbit and parity ($p$-, $d$-orbitals) degrees of freedom in centrosymmetric transition metal dichacogenides , which is called as spin-orbit-parity coupled (SOPC) superconductors. In this work, we show that  the SOPC would induce a strong spin-singlet and spin-triplet pairing mixing near the topological band inversion in this system. Moreover, we find that the presence of such mixing provides a close explanation of the observed the in-plane upper critical fields $B_{c2}$ in terms of both the enhancement and  anisotropy. We also propose measuring the equal-spin Andreev reflection between 2M-WS$_2$ and a ferromagnetic (FM) lead to detect the spin-triplet pairing in 2M-WS$_2$. Our work paves a way to study the spin-singlet and spin-triplet pairing in centrosymmetric superconductors with strong spin-orbital parity coupling.
\end{abstract}


\maketitle
{\emph {Introduction.}}---
A superconductor can be classified into noncentrosymmetric and centrosymmetric superconductors according to its crystal symmetry. In a noncentrosymmetric superconductor,  the spin-singlet and spin-triplet pairing would generally mix with each other in the  presence of strong spin-orbit coupling $\bm{g}(\bm{k})\cdot \bm{\sigma}$ ($\bm{k}$ is momentum, $\sigma_j$ is Pauli matrices)  \cite{Rashba2001}. One important consequence of  this paring mixing is to enhance the upper critical fields of noncentrosymmetric superconductors \cite{Smidman_2017,bauer2012non}, which have been well explored in some noncentrosymmetric superconductors, such as heavy fermion superconductors \cite{Sigrist2004}, Ising superconductor \cite{Lu2015, Xi2016, Saito2016, delaBarrera2018, Ye2018, Xing2017, Sohn2018, Tongzhou2016, He2018, Yingming2020,Mazin2020}. For centrosymemtric superconductors, due to the combination of both time-reveral symmetry and inversion symmetry, the usual spin strong spin-orbit coupling $\bm{g}(\bm{k})\cdot \bm{\sigma}$ vanishes. As a result, the spin-singlet and spin-triplet pairing mixing are often overlooked in centrosymmetric superconductors.

In recent years, several centrosymmetric superconducting thin films  with in-plane upper critical fields  beyond Pauli limit ($B_p$) are seen in the experiments \cite{Fatemi,Sajadi,Faxian2022,Leslie2022,Liu2020,XueQi2020}. Notably, a large class of them  are represented by various centrosymmetric transition metal dichaneides thin film, the monolayer of which are predicted to be a two-dimensional quantum spin Hall insulator \cite{Qian,roberto2016,Tang2017,Fei2017,WuSanfeng}, including 1T$'$-WTe$_2$ \cite{Fatemi,Sajadi}, 2M-WS2 \cite{Faxian2022}, 1T$'$-WS2\cite{Leslie2022}. Specifically, in the year 2018, centrosymmetric monolayer 1T$'$-WTe$_2$ was reported to be superconducting upon electrograting while the in-plane $B_{c2}$ is one to three times $B_p$ \cite{Fatemi, Sajadi}. Due to  the presence of inversion symmetry, the spin-orbit coupling term that involves only spin and momentum, which is widely used to explain such enhancement in noncentrosymmetric superconductors,  is not allowed in 1T$'$-WTe$_2$. Hence, another intrinsic mechanism behind the observed enhance-ment of $B_{c2}$ was pointed out in Ref.~\cite{Law2020}. It shows that the spin-orbit-party coupling that involves the coupling the spin, orbit and party degrees of freedom in this system would renormalize the spin susceptibility and  an anisotropic $B_{c2}$ higher than the Pauli limit is predicted.  This prediction  is  clearly demonstrated in the centrosymmetric superconducting 2M-WS2 thin film recently \cite{Faxian2022}, which exhibits the same structure 1T$'$-structure in monolayer but displays a distinct stacking along $z$-axis. Notably, similar anisotropically enhanced $B_{c2}$ beyond $B_p$ has also been seen in newly fabricated centrosymmetric 1T$'$-WS2 thin film \cite{Leslie2022}, which could be also explained in terms of SOPC. Given these experimental progresses, a study of whether the SOPC would induce spin-singlet and spin-triplet pairing mixing  in these centrosymmetric superconductors would be highly desirable. Also, how this pairing mixing would affect the $B_{c2}$ is also an interesting question.

In this work, using a model that captures the realistic bands of centrosymmetric TMD, we study the spin-singlet and spin-triplet pairing mixing induced by the strong SOPC in centrosymmetric superconductors. We first show that due to the presence of SOPC, the inter-orbital spin triplet pairing correlation would be non-zero. Next, we classify the possible pairings according to the irreducible representation of the point group symmetry. Importantly, we find that the trivial $A_g$ representation contains both intra-orbital spin-singlet and inter-orbital spin-triplet pairing. As a result, a mixing between them is possible. Then, we calculate the superconducting phase diagram involving these two pairings, which shows a clear mixing region. Moreover, by calculating the superconducting pairing susceptibility at different SOPC strength, we find that the mixing strength is enhanced with the  SOPC strength. Finally, we discuss  the possible connection to the experiment by showing the in-plane $B_{c2}$ with and without spin-singlet and spin-triplet pairing mixing. We find the pairing mixing helps to enhance both the magnitude  and anisotropy of the $B_{c2}$, which provides a close fitting of the $B_{c2}$ measured by the experiment. We also find that when a ferromagnetic lead is attached to 2M-WS$_2$, the tunneling amplitude is anisotropic due to equal-spin Andreev reflection, suggesting the existence of spin-triplet pairing in 2M-WS$_2$.

\begin{figure}
	\centering
	\includegraphics[width=1\linewidth]{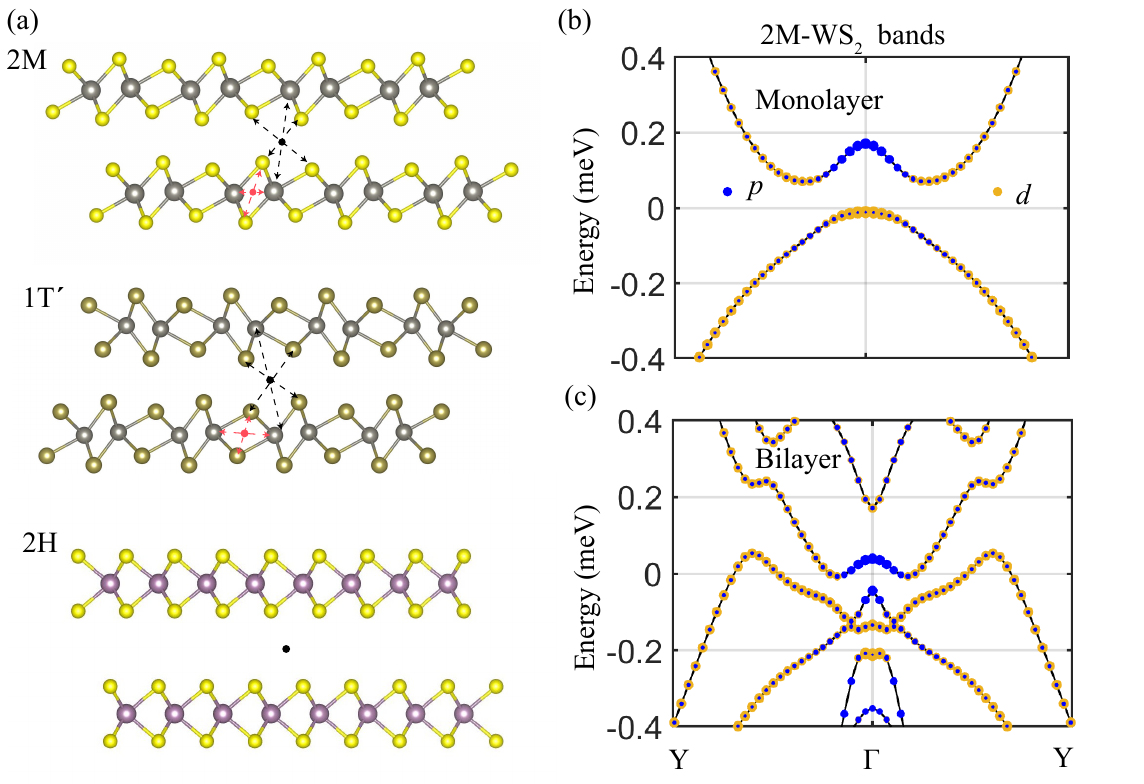}
	\caption{(a) shows   bilayer transition metal dichagenides with 2M-, 1T$'$-, 2H-structures, respectively. The  bilayer inversion center and monolayer layer inversion  center are highlighted as black and red dots, respectively. (b) and (c) show the first-principle calculated band structures of a monolayer and bilayer 2M-WS$_2$, respectively. The projected $d$ states for the tungsten atoms and $p$ states for the sulfur atoms 
	are shown.  }
	\label{fig:fig1}
\end{figure}

{\emph {Model.}}---
We first present a model which describes a centrosymemtric superconductor with strong SOPC. This model is motivated by the recent finding of superconductivity in centrosymmeric superconducting 2M-WS$_2$, 1T$'$-WTe$_2$ thin film. An illustration of these crystal structures is depicted in Fig.~\ref{fig:fig1}(a), which shows a bilayer structure of them. It can be seen that both the  monolayer and bilayer 1T$'$- and 2M-structure are centrosymmetric (the inversion center are highlighted as red and back dot, respectively ).  It is distinct from the notable 2H-structure, in which although even layers are centrosymmetric but is always locally noncentrosymmetric within each layer (see bottom panel of Fig.~\ref{fig:fig1}(a)). 

Let us first model the monolayer, which is identical for 2M- and 1T$'$-structure and the bands are topological \cite{Qian}.
 The states near Fermi energy in this case are contributed by a topological band inversion between a $p$-orbital dominant band and a $d$-orbital dominant band. As an illustration of this topological band inversion, the monolayer 2M-WS$_2$ energy bands obtained from first principle calculation is shown in Fig.~\ref{fig:fig1}(b). To describe these bands, a four-band  $\bm{k}\cdot \bm{p}$ Hamiltonian  dedicated by the point symmetry $C_{2h}$ to describe normal states is written as \cite{Faxian2022, Law2020}
   \begin{eqnarray}
   H_0(\mathbf{k})=&&\epsilon_{\mathbf{k}}+\mathcal{M}_{\mathbf{k}}s_z
  	+vk_xs_y+(A_xk_xs_x\sigma_y\nonumber\\&&+A_yk_ys_x\sigma_x
  	+A_zk_ys_x\sigma_z)+\frac{1}{2}g_su_B\bm{B}\cdot\bm{\sigma},\label{Eq_mono}
  \end{eqnarray}
  Here,  $s$ are Pauli matrices operating on ($p,d$) orbitals, $\sigma$ are Pauli matrices operating on spin space. $\epsilon_{\mathbf{k}}=t_{0x}k_x^2+t_{0y}k_y^2+t_{2x}k_x^4+t_{2y}k_y^4-\mu$, $M(\bm{k})=-\delta+m_{0x}k_x^2+m_{0y}k_y^2+m_{2x}k_x^4+m_{2y}k_y^4$, where $\mu$ is the chemical potential. The SOPC term in the Hamiltonian is represented by $(A_xk_x\sigma_y, A_yk_y\sigma_x,A_zk_y\sigma_z)s_x$. It would be expected that the SOPC is strongest near the topological band inversion due to $s_x$, which involves the mixing of two orbitals. The last term is the Zeeman term, where $\bm{B}$ is set along in-plane directions and $g$-factor is set to be 2.
  
  In a thin-film structure, the interlayer coupling would make the system to become metallic even without doping. As an illustration, we plots the energy bands of the bilayer 2M-WS2 in Fig.~\ref{fig:fig1}(c). It should be noted that the SOPC arising from the $p$-, $d$-band inversion and strong spin-orbit coupling  still exhibits in this case. 
  We also construct a   $\bm{k}\cdot \bm{p}$ Hamiltonian of a  bilayer 2M-WS2 by coupling two copies of monolayer Hamiltonian with interlayer couplings. The detailed form of this bilayer Hamiltonian and model parameters obtained from fitting the first principle calculated bands are listed in Supplementary Sec.~I. 
  
In the superconducting state, we can describe the system with Bogoliubov-de Gennes (BdG) Hamiltonian:
   \begin{equation}
  	H_{BdG}(\mathbf{k})=\begin{pmatrix}
  		H_0(\mathbf{k})& \Delta(\bm{k}) i\sigma_y\\
  		(\Delta (\bm{k}) i\sigma_y)^{\dagger}& -H_0^{*}(-\mathbf{k})
  	\end{pmatrix}.
  \end{equation}
where $H_0(\bm{k})$ is the normal state Hamiltonian, $\Delta(\bm{k})$ denotes the superconducting order parameters. Later, we will classify all possible pairings according to the irreducible representations of $C_{2h}$.
   
\emph{Spin-triplet correlations.}--- Let us first consider the simplest case with $\Delta(\bm{k})=\Delta$ and $H_0(\bm{k})$ takes the form of Eq.~(\ref{Eq_mono}). We would show that due to the presence of strong SOPC, the spin-triplet pairing correlation in Green's function is finite even with this  uniform s-wave pairing order parameter.
 
The Green's function is a convenient way to capture the superconducting properties, which can be expressed with
\begin{eqnarray}
 	G_{\lambda\mu}(\bm{k},\tau)&&=-\braket{T_{\tau}\{c_{\bm{k},\lambda}(\tau)c^{\dagger}_{\bm{k},\mu}(0)\}},\\
 	F_{\lambda\mu}(\bm{k},\tau)&&=\braket{T_{\tau}\{c_{\bm{k},\lambda}(\tau)c_{-\bm{k},\mu}(0)\}}.
\end{eqnarray}
Here, $\tau$ is imginary time, $\lambda,\mu$ are internal degrees of freedom, $T_{\tau}$ is the time-ordering operator.  We can rewrite the Green's function in the Matsubara frequency space: $G_{\lambda\mu}(\bm{k},i\omega_n)=\int_0^{\beta}d\tau e^{i\omega_n\tau}G_{\lambda\mu}(\bm{k},\tau)$ and $F_{\lambda\mu}(\bm{k},i\omega_n)=\int_0^{\beta}d\tau e^{i\omega_n\tau}F_{\lambda\mu}(\bm{k},\tau)$. The latter $F_{\lambda\mu}(\bm{k},i\omega_n)$ represents the pairing correlations we refer, which captures the properties of Cooper pairs.  These two Green's functions are related to the Gor'kov Green's function as
\begin{equation}
 	\mathcal{G}(\bm{k},i
 	\omega_n)=\begin{pmatrix}
 		G(\bm{k},i\omega_n)&-F(\bm{k},i\omega_n)\\-F^{\dagger}(\bm{k},i\omega_n)&-G^{T}(-\bm{k},-i\omega_n)
 	\end{pmatrix}.
\end{equation}
The Gor'kov Green's function can be calculated as $\mathcal{G}(\bm{k},i
 	\omega_n) = (i\omega_n-H_{BdG}(\bm{k}))^{-1}$. 
Insert the BdG Hamiltonian and after some massage, we can parameterize the pairing correlation as 
\begin{equation}
    F(\bm{k},i\omega_n)=\Delta[C_1(\bm{k},i\omega_n) + C_2(\bm{k},i\omega_n)s_x\bm{d}(\bm{k})\cdot \bm{\sigma}] i\sigma_y\label{pairing _corr},
\end{equation}
where the detailed expressions of coefficients $C_1(\bm{k},i\omega_n)$ and $C_2(\bm{k},i\omega_n)$ are presented in Supplementary Sec. II. 

Notably, it can be seen that there is a triplet component in the pairing correlation, where we find that the triplet $d$-vector is the SOPC vector (see Supplementary Sec.~II): 
 \begin{equation}
	\bm{d}(\bm{k})\approx (A_yk_y,A_xk_x,A_zk_y). \label{eq:d_vector}
\end{equation}
The structure of the correlation $F(\bm{k},i\omega_n)$ indicates that due to the presence of SOPC, the intra-orbital spin-singlet pairing and the interorbbital spin-triplet pairing are mixed with each other. 

\begin{table}
\caption{Classifications of possible time-reversal-invariant intralayer pairings according to the irreducible representations (IRs) of $C_{2h}$ point group for a bilayer 2M-WS$_2$. Here the triplet $\bm{d}$-vector is along the direction of SOPC, i.e., 	$\bm{d}(\bm{k})= (A_yk_y,A_xk_x,A_zk_y)$ and $\bm{\hat{d}}(\bm{k})=\bm{d}(\bm{k})/|\bm{d}(\bm{k})|$.  }
    \label{symmetry_classfication}
	\begin{tabular}{cccc}
		\hline\hline
		IRs &\hspace{1 mm} $A_g$  &\hspace{1 mm} $A_u$ &\hspace{1 mm} $B_u$\\\hline	
		$I=s_z$ &\hspace{1 mm} $+$ &\hspace{1 mm} $-$ &\hspace{1 mm} $-$\\
		$C_{2y}=is_z\sigma_y$ &\hspace{1 mm} $+$ &\hspace{1 mm} $+$ &\hspace{1 mm} $-$\\\hline
		Spin-singlet &\hspace{1 mm} $s_0$, $s_z$ &\hspace{1 mm} None &\hspace{1 mm} $ s_x$\\
		Spin-triplet  &\hspace{1 mm}   $s_x \bm{\hat{d}}(\bm{k})\cdot \bm{\sigma}$          &\hspace{1 mm} $ s_y\sigma_x$, $ s_y\sigma_z$ &\hspace{1 mm} $ s_y\sigma_y$, $\bm{\hat{d}}(\bm{k})\cdot \bm{\sigma}$\\\hline
		\hline
	\end{tabular}
	\label{TableS3}
\end{table}
\emph{Symmetry analysis.}--From the pairing correlation $F(\bm{k},i\omega)$ in Eq.~(\ref{pairing _corr}), we can see that the intra-orbital spin-singlet pairing and the inter-orbital spin-triplet pairing in general would mix with each other. Next, let us show this from the symmetry point of view. The pairing order parameter can be classified according to the transformation properties of the paring matrix $\Delta(\bm{k})$ under the group generators of $C_{2h}$ including an inversion operator $I=s_z$ and a two-fold rotation $C_{2y}=is_z\sigma_y$. 

We summarize all possible time-reversal-invariant intralayer pairings $C_{2h}$ in Table~\ref{symmetry_classfication}
according to the irreducible representations of $C_{2h}$.  The nontrivial $A_u$-pairing  includes two  inter-orbital spin-triplet pairings $\Delta_{A_{u,1}}=s_y\sigma_x$ and $\Delta_{A_{u,2}}=s_y\sigma_z$. The nontrivial $B_u$-pairing includes a spin-singlet pair $\Delta_{B_{u,1}}=s_x$ and triplet pairings $\Delta_{B_{u,2}}=s_y\sigma_y$ and $\Delta_{B_{u,3}}=\bm{\hat{d}}(\bm{k})\cdot\sigma_y$. As we found all the nontrivial $A_{u}$- and $B_{u}$-pairings lead to a divergent $B_{c2}$ when $T\rightarrow 0$, which are unlikely in the experiment. We would thus  focus on $A_{g}$-pairing only. 

The $A_g$-pairing allows an intra-orbital spin-singlet pairing, and particularly also allows an inter-orbital spin-singlet pairing $\Delta_{Ag,2}=s_x\bm{\hat{d}}(\bm{k})\cdot \bm{\sigma}$. Similar to the noncentrosymemtryic superconductors case \cite{Sigrist2004}, the triplet $\bm{\hat{d}}(\bm{k})$-vector should be parallel to the spin-orbit coupling vector in order to save free energy. As these two pairings belong to the same irreducible representation, they generally would mix with each other, being consistent with the pairing correlations shown previously (Eq.~(\ref{pairing _corr})). Our analysis clearly shows the discussion of spin-singlet and spin-triplet pairing proposed in noncentrosymmetric superconductors \cite{Rashba2001,Smidman_2017,Sigrist2004,bauer2012non} can be extended even in centrosymmetric superconductors.  

\emph{Superconducting phase diagrams and pairing susceptibility.}--- To show how the mixing strength is affected by interaction strength and the SOPC strength, we study the superconducting phase diagram of $A_g$ pairing and show how these two pairings are stabilized under various SOPC strength.  
 
 The superconducting phase diagram can be obtained by solving the linearized gap equation. For simplicity, we negelect $s_z$ in $A_{g,1}$ , which only represents a pairing gap difference between two orbitals. Now we study the pairing instability of $A_g$ representation with a conventional intra-orbital spin-singlet pairing $\Delta_{A_{g,1}}=s_0$ and an inter-orbital spin-triplet pairing $ \Delta_{A_{g,2}}\equiv\bm{d}_{A_{g,2}}=s_x\bm{\hat{d}}\cdot\bm{\sigma}$. The corresponding linearized equation is written as 
 \begin{equation}
 	\text{det}\begin{bmatrix}
 		\begin{pmatrix}
 			\frac{1}{U}-\chi^{p}_{ 11}(T_c)&-\chi^{p}_{12}(T_c)\\
 			-\chi^{p}_{ 21}(T_c)&\frac{1}{V}-\chi^{p}_{22}(T_c)
 		\end{pmatrix}
 	\end{bmatrix}=0,\label{eq:pairingeq}
 \end{equation}
 where $\chi^p_{mm'}$ denotes the pairing susceptibility:
$\chi^p_{mm'}=-\frac{1}{\beta}\sum_{n,\bm{k}}\text{Tr}[G_e(\bm{k},i\omega_n)(\Delta_{\Gamma,m}i\sigma_y)G_h(\bm{k},i\omega_n)(\Delta_{\Gamma,m'}i \sigma_y)^{\dagger}]
$. Here, the single particle electron Green's function $G_e(\bm{k},i\omega_n)=(i\omega_n-H_0(\bm{k}))^{-1}$ and hole Green's function $G_h(\bm{k},i\omega_n)=(i\omega_n+H^*_0(-\bm{k}))^{-1}$, where $H_0(\bm{k})$ is the normal  Hamiltonian. The pairing susceptibility can be calculated numerically in band basis. We label the intra-orbital interaction strength  and inter-orbital interaction strength  as $U$ and $V$ respectively. 
 
 \begin{figure}
 	\centering
 	\includegraphics[width=1\linewidth]{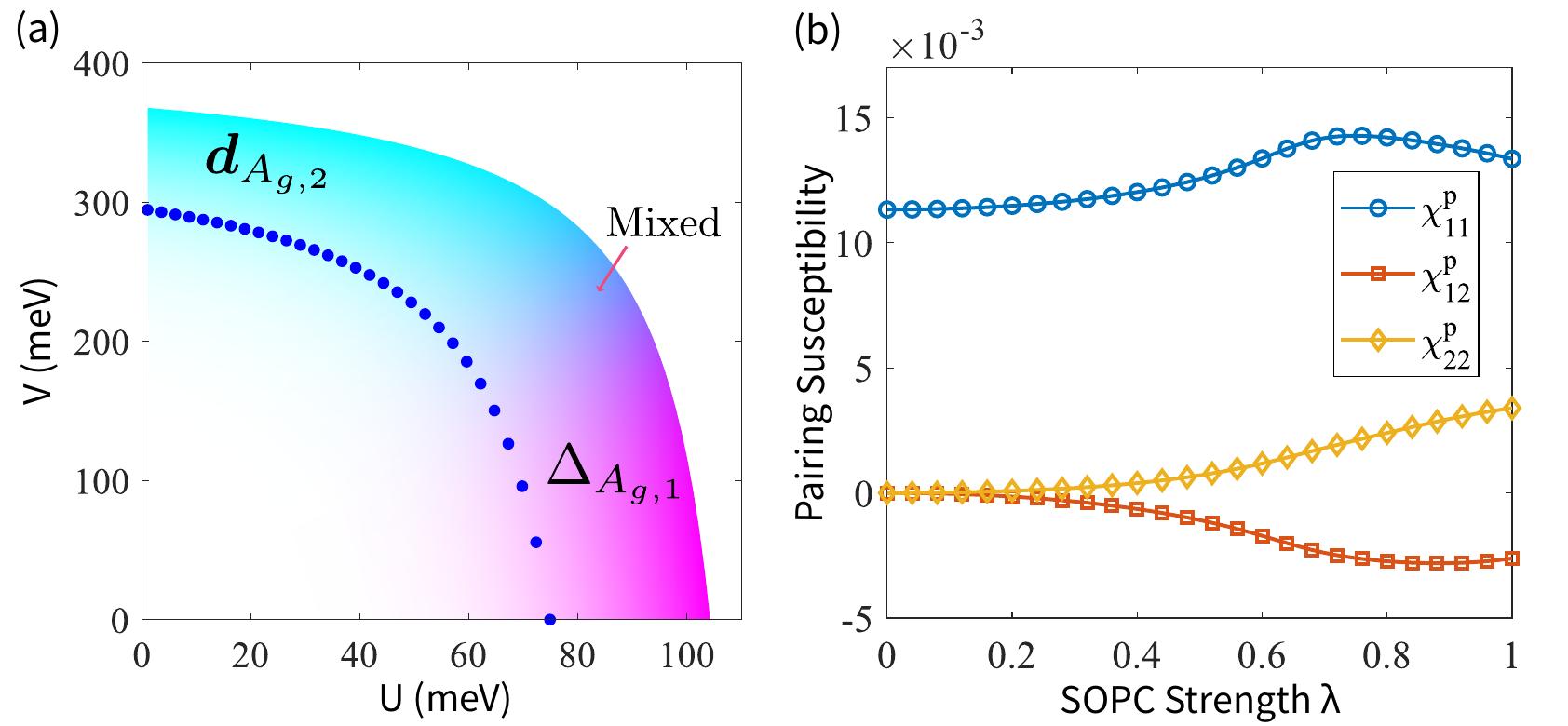}
 	\caption{(a) shows the superconducting phase diagram with intra-orbital spin-singlet pairing $\Delta_{A_{g,1}}$ and  inter-orbital spin-triplet pairing $\bm{d}_{A_{g2}}$. The blue dots denote the regime with critical temperature of 7.6 K. (b) shows the  pairing susceptibility of  $\chi^{p}$  as a function of SOPC strength $\lambda$, where  $\chi_{11}^{p}$ and $\chi_{22}^{p}$ represent  the  pairing susceptibility of the pairing $\Delta_{A_{g,1}}$ and $\bm{d}_{A_{g2}}$, respectively.  $\chi_{12}^{p}$ represents the mixed pairing susceptibility between the pairing $\Delta_{A_{g,1}}$ and $\bm{d}_{A_{g2}}$. }
 	\label{fig:fig2}
 \end{figure}
 
Figure \ref{fig:fig2}(a) shows the calculated superconducting phase diagram with intra-orbital spin-singlet pairing $\Delta_{A_{g,1}}$ and  inter-orbital spin-triplet pairing $\bm{d}_{A_{g2}}$. The critical temperature is determined by solving Eq.~(\ref{eq:pairingeq}) and the mixing (represented by the color in Fig.~\ref{fig:fig2}(a)) is obtained from the eigenvectors of Eq.~(\ref{eq:pairingeq}). As expected, the conventional intra-orbital spin-singlet $A_{g,1}$ pairing is more favorable when the intra-orbital interaction strength $U$ is stronger, while  the inter-orbital spin-triplet $A_{g,2}$ pairing is more favorable when the inter-orbital interaction strength $V$ is stronger. At certain $(U,V)$ regime, a prominent mixed region can also be seen, as highlighted in Fig.~\ref{fig:fig2}(a). This mixing is allowed by symmetries since both pairings belong to $A_g$ representation. As discussed in the pairing correlations (Eq.~\ref{pairing _corr}), such mixing is induced by the SOPC. A purely $\bm{d}_{A_{g2}}$ pairing region can also be seen in Fig.~\ref{fig:fig2}(a).
 
In Fig.~\ref{fig:fig2}(b), we display the  pairing susceptibility of $\chi^{p}$  as a function of SOPC strength $\lambda$, where the SOPC coefficients are scaled as $\lambda A_j$. It can  be seen that the mixing vanishes ($\chi^{p}_{12}=0$) when there is no SOPC ($\lambda=0$). We also noted the magnitude of the pairing susceptibility $\chi_{12}^p$ increases with the SOPC strength. It means that the spin-triplet pairing $\bm{d}_{A_{g2}}$ is  more favorable when the SOPC is stronger. The enhancement of $\bm{d}_{A_{g2}}$ pairing stems from the fact that  $\bm{d}_{A_{g2}}$ pairing is inter-orbital pairing and the SOPC can introduce the orbital mixing. In contrast, as shown in Fig.~\ref{fig:fig2}(b), the pairing susceptibility of conventional spin-singlet pairing $\Delta_{A_{g,1}}$ is relatively insensitive to the SOPC strength.
\begin{figure}
	\centering
	\includegraphics[width=1\linewidth]{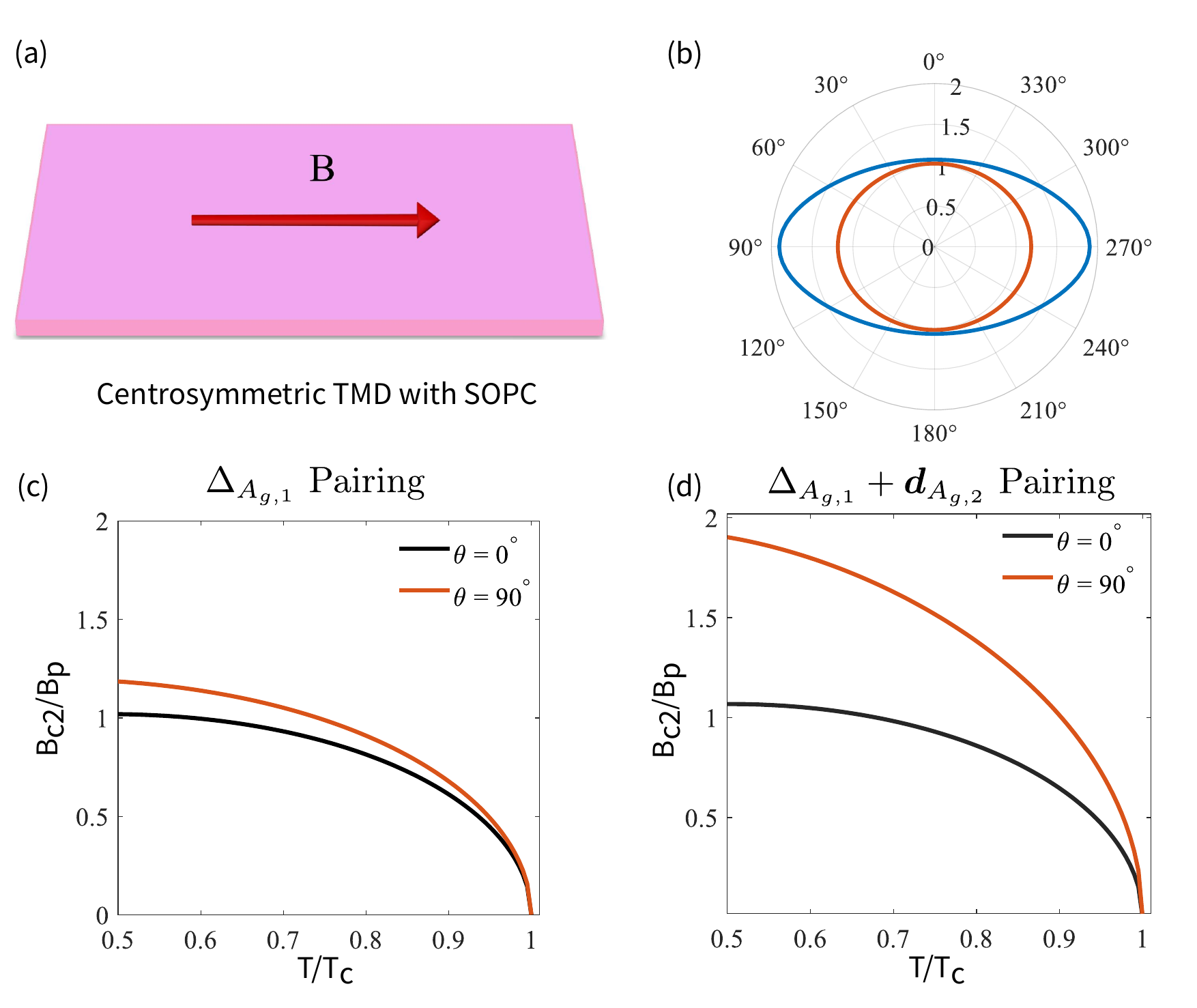}
	\caption{(a) schematically shows the a centrosymmetric TMD, such as the 2M-WS$_2$ thin film, in the presence of in-plane fields. (b) shows the angular dependence of in-plane upper critical fields for a bilayer 2M-WS$_2$ with purely $\Delta_{A_{g1}}$ pairing (red) and  $\Delta_{A_{g1}}+\bm{d}_{A_{g2}}$ mixed pairing (blue), where the radius shows the value of $B_{c2}/B_p$, and the temperature is fixed at $T=0.5 T_c$. (c) and (d) The calculated  $B_{c2}/B_p$ versus $T/T_c$ curve for the 2M-WS$_2$ with the in-plane magnetic field
	direction applied along $\theta=0^{\circ}$ and $\theta=90^{\circ}$, respectively. Here, we have adopted the interaction parameters $(U,V)=(75,0)$ meV for $\Delta_{A_{g1}}$ pairing and $(U,V)=(10.7,288)$ meV for $\Delta_{A_{g1}}+\bm{d}_{A_{g2}}$ pairing.}
	\label{fig:fig3}
\end{figure}
 
 \emph{In-plane upper critical fields.} As we discussed, the symmetry allows a mixed of $\Delta_{A_{g1}}$ pairing and $d_{A_{g2}}$ pairing. In this section,  we show that how the in-plane upper critical fields (Fig.~\ref{fig:fig3}(a)) are changed in the presence of the additional $\bm{d}_{A_{g2}}$ pairing mixing.  In this case,   we need to solve the linearized gap equation Eq.~(\ref{eq:pairingeq}) with replacing the pairing susceptibilities $\chi^{p}(T)$ with ones of finite fields  $\chi^{p}(B,T)$. i.e.,
 \begin{equation}
 	\text{det}\begin{bmatrix}
 		\begin{pmatrix}
 			\frac{1}{U}-\chi^{p}_{ 11}(B,T)&-\chi^{p}_{12}(B,T)\\
 			-\chi^{p}_{ 21}(B,T)&\frac{1}{V}-\chi^{p}_{22}(B,T)
 		\end{pmatrix}
 	\end{bmatrix}=0.\label{eq:finite_B}
 \end{equation}
 
The results are summarized in Fig.~\ref{fig:fig3}.  Fig.~\ref{fig:fig3}(b) shows the angular dependence of $B_{c2}/B_{p}$ of a purely spin-singlet state $\Delta_{A_{g1}}$ and  spin-singlet (blue) and spin-triplet mixed state (red). It clearly shows that the enhancement and anisotropic ratio of $B_{c2}$ can be enhanced by the mixing of $\bm{d}_{A_{g2}}$ pairing. In Fig.~\ref{fig:fig3}(c) and Fig.~\ref{fig:fig3}(d), the calculated in-plane critical magnetic fields  as a function of temperature are plotted for a purely $\Delta_{A_{g1}}$ pairing  and $\Delta_{A_{g1}}+\bm{d}_{A_{g2}}$. Consistently, we find the enhancement for the $\Delta_{A_{g1}}+\bm{d}_{A_{g2}}$ pairing is larger than a purely $\Delta_{A_{g1}}$ pairing. More importantly, the sizable enhancement  $B_{c2}\approx 2 B_p$ for fields along $\theta=90^\circ$ suggested by $\Delta_{A_{g1}}+\bm{d}_{A_{g2}}$ pairing is very close to the experimental data of 2M-WS2 thin-film \cite{Faxian2022}. 
 
The $\bm{d}_{A_{g2}}$-pairing enhances the critical magnetic fields is understandable, because it is a spin-triplet pairing which can induce equal-spin cooper pairs and save magentic energy in a superconducting state. However, this spin-triplet pairing would not lead to a divergent $B$-fields as $T\rightarrow0$ as a usual spin-triplet pairing even in the absence of orbital effects. This is because the direction of spin-triplet  $\bm{d}$-vector is not fixed within the momentum space for  this centrosymmetric system. It is significantly differnt from 2H-type superconducting TMDs, where the favorable spin-triplet $\bm{d}$-vector is pinned along out of plane direction \cite{Tongzhou2016,  Yingming2020}. 

\emph{Andreev reflection}--- To further verify our theory, we propose to attach a ferromagnetic lead to these centrosymmetric superconducting TMDs and calculate for the Andreev reflection. 

We model the ferromagnetic lead with a square lattice. For simplicity, we use a monolayer $\bm{k\cdot p}$ Hamiltonian in tight-binding limit. Using the Fisher-Lee trick \cite{fisherlee1, fisherlee2}, the scattering matrix is given by
\begin{equation}
    S = \begin{pmatrix}
        r_{ee} & r_{eh}\\
        r_{he} & r_{hh}
    \end{pmatrix} 
    = -I +i\Gamma^{1/2}G^R(E)\Gamma^{1/2},
\end{equation}
where $r_{ee}, r_{he}$ are the electron and Andreev reflection matrix respectively. $\Gamma = i(\Sigma-\Sigma^\dagger)$ is the broadening function, $\Sigma$ is the self-energy, and $G^R(E)$ is the total (retarded) Green's function. The conductance is then given by (c.f.~\cite{Manna2020})
\begin{equation}
    G(E) = \frac{e^2}{h} \text{Tr}\left[I-r_{ee}^\dagger r_{ee}+r_{he}^\dagger r_{he}\right].
\end{equation}
Fig.~\ref{fig:fig4}(b)-(d) show the in-gap conductance at different magnetization amplitude. Similar to Fig.~\ref{fig:fig3}, when rotating the polarization direction of the ferromagnetic lead, the in-gap Andreev reflection exhibits anisotropy with a two-fold symmetry, showing suppression at angles of $\theta = 0^\circ, 180^\circ$. This is due to the triplet pairing correlations induced by the strong SOPC. In 2M-WS$_2$, the triplet $\bm{d}$-vector at the Fermi surface is approximately parallel to the $x$-axis. This favours equal-spin triplet pairing in $y$-direction. As a result, when the magnetization angle is at $0^\circ$, Andreev reflection of equal-spin holes is suppressed \cite{Liu2020}. In contrast, approaching to $90^\circ$, equal-spin hole can be reflected. The same argument repeats for $180^\circ$ and $270^\circ$. The results should be similar in other SOPC superconductors since it is mostly controlled by the triplet $d$-vector which is proportional to the SOPC vector. We also include a similar calculation for 1T$'$-WTe$_2$ in Supplementary Sec.~IV.

\begin{figure}
	\centering
	\includegraphics[width=1\linewidth]{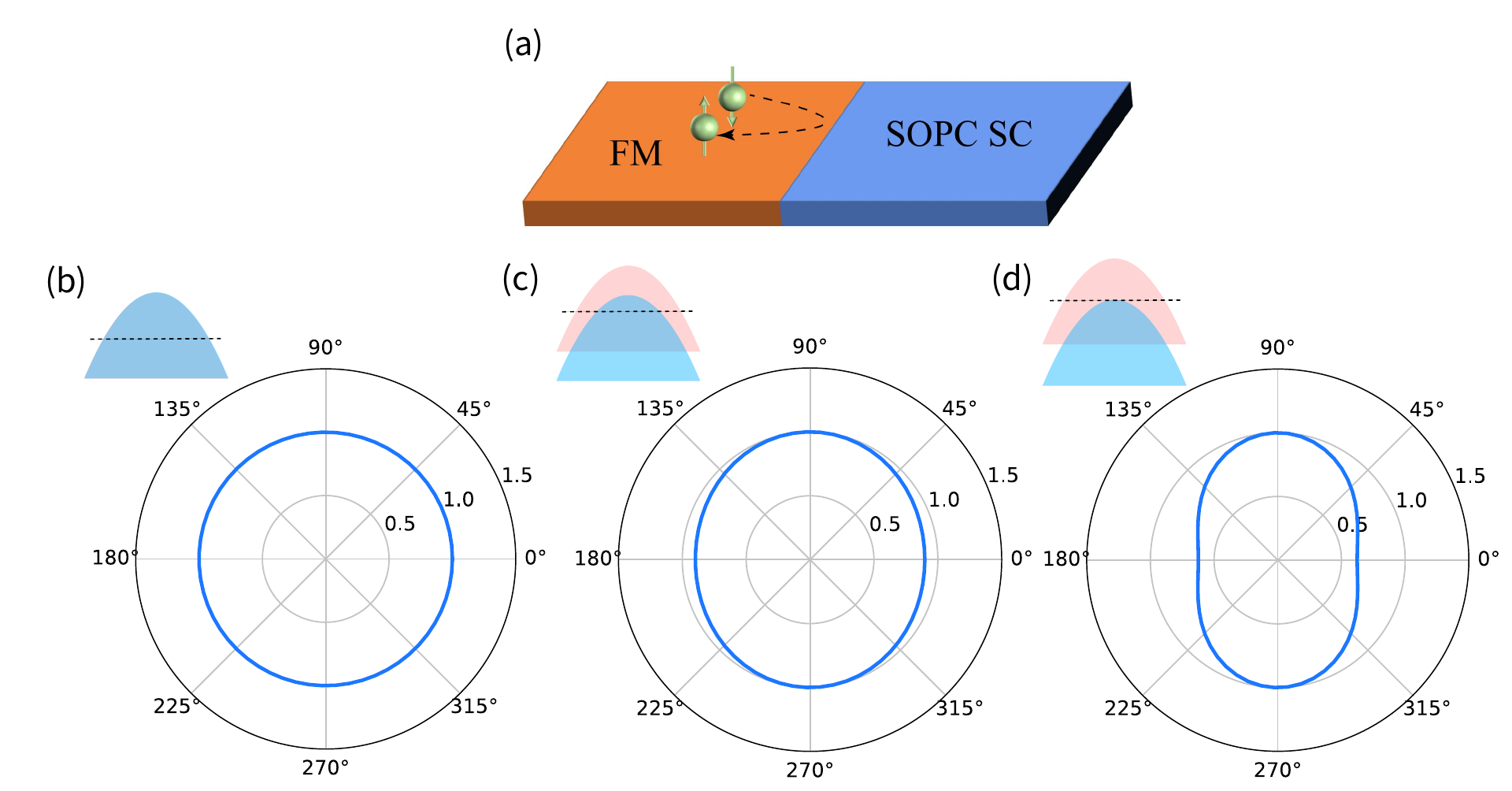}
	\caption{(a) The schematic plot of a ferromagnet metal/SOPC superconductor (SC) tunneling junction. The Andreev reflection process is highlighted on the plot. (b)-(d) The angular dependence of in-gap normalized tunnel conductance $G$, where the angle represents the in-plane magnetization direction. From left to right, we . The major spin and minor spin population are illustrated in each case. It can be clearly seen that the Anreev reflection amplitude exhibits a two-fold feature when $m$ is finite in the ferromagnetic metal lead. The two-fold anisotropic feature will become more salient when ferromagentic metal lead is close to a half-metal in, in which almost only half of the spin population is occupied.  }
	\label{fig:fig4}
\end{figure}
 
 \emph{Conclusions.}---In conclusion, we have established a self-consistent mean-field theory to demonstrate singlet-triplet mixed stated induced by the SOPC in centrosymmetric superconductors.
 In particular, our finding explains the observed enhancement of $B_{c2}$ with a large anisotropic ratio in 2M-WS$_2$ thin film.
 Furthermore, we find anisotropic in-gap Andreev reflection amplitudes at the FM/SOPC superconductor tunneling junction with respect to the magnetization direction of the FM lead. The result can be experimentally tested to verify spin-triplet pairing in general SOPC superconductor. 
 Our theory generally applies to other similar centrosymmetric superconducting TMDs, such as 1T$'$-WTe$_2$, 1T$'$-WS$_2$.
 
 {\emph { Acknowledgments.}}--- Tsz Fung Heung and Yip Chun Wong sincerely express their heartfelt appreciation for the invaluable insights and guidance offered by Yingming Xie, Benjamin T. Zhou and K.T. Law. They also extend their sincere gratitude for the DFT calculation data of 2M-WS$_2$ provided by Xue-Jian Gao.

\clearpage
\onecolumngrid
\begin{center}
		\textbf{\large Supplementary Material for\\ ``Spin-orbital-parity coupling induced singlet-triplet mixed state  in  centrosymmetric superconducting  transition metal dichagenides''}\\[.2cm]		
      Tsz Fung Heung,$^{*,1}$ Yip Chun Wong,$^{*,1}$\\[.1cm]
		{\itshape ${}^1$Department of Physics, Hong Kong University of Science and Technology, Clear Water Water Bay,  Hong Kong, China}
\end{center}
\maketitle

\setcounter{equation}{0}
\setcounter{figure}{0}
\setcounter{table}{0}
\setcounter{page}{1}
\setcounter{section}{0}
\setcounter{secnumdepth}{4}
\makeatletter
\renewcommand{\theequation}{S\arabic{equation}}
\renewcommand{\thefigure}{S\arabic{figure}}
\renewcommand{\thetable}{S\arabic{table}} 
\renewcommand{\thesection}{\Roman{section}} 

\renewcommand{\theHtable}{Supplement.\thetable}
\renewcommand{\theHfigure}{Supplement.\thefigure}
\renewcommand{\theHequation}{Supplement.\theequation}
\renewcommand{\theHsection}{Supplement.\thesection}

\renewcommand{\bibnumfmt}[1]{[S#1]}
\renewcommand{\citenumfont}[1]{S#1}
\maketitle

\section{Effective $\mathbf{k}\cdot \mathbf{p}$ Hamiltonians }

The crystal structure of 2M-WS$_2$ respects to point group $C_{2h}$. The generators are an inversion symmetry and a two-fold symmetry along $y$ axis. From the results of first principle calculation, the bands near Fermi energy are dominant by the orbitals $\ket{S,A_u}$, $\ket{W, B_g}$. Here, $A_u$ corresponds to a $p$-wave-like orbit, while $B_g$ corresponds to a $d$-wave-like orbit. As a result, the representations of the generators can be written as $C_{2y}=i\sigma_ys_z\tau_x$ and $I=s_z\tau_x$, where Pauli matrices  $\tau$, $s$, $\sigma$ operate on layer, orbital, spin space respectively. 

Based on these symmetry generators, a $\mathbf{k\cdot p}$ normal Hamiltonian for monolayer 2M-WS$_2$ dictated by the $C_{2h}$ symmetry is obtained as
\begin{equation}\label{eq:monolayer_kdotp}
	H_0(\mathbf{k})=\epsilon(\mathbf{k})+\mathcal{M}(\mathbf{k})s_z+v(\bm{k})k_xs_y+(A_xk_xs_x\sigma_y+A_yk_ys_x\sigma_x+A_zk_ys_x\sigma_z).
\end{equation}
where $\epsilon(\mathbf{k})=t_{0x}k_x^2+t_{0y}k_y^2+t_{2x}k_x^4+t_{2y}k_y^4-\mu$, $\mathcal{M}(\mathbf{k})=-\delta+m_{0x}k_x^2+m_{0y}k_y^2+m_{2x}k_x^4+m_{2y}k_y^4$, $v(\bm{k})=v_1+v_2k_x^2+v_3k_y^2$. $\mu$ is the chemical potential.

By adding the interlayer coupling Hamiltonian, we can construct an effective Hamiltonian for a bilayer 2M-WS2:
\begin{equation}\label{eq:bilayer_kdotp}
	H_N(\mathbf{k})=H_0(\bm{k})\tau_0 + H_c(\bm{k})\sigma_0
\end{equation}
Here the basis $\psi_{\mathbf{k}}=(\psi_{\mathbf{k},+},\psi_{\mathbf{k},-})^{T}$ with  $\psi_{\mathbf{k},l}=(c_{\mathbf{k},p,\uparrow,l},c_{\mathbf{k},p,\downarrow}, c_{\mathbf{k},d,\uparrow,l},c_{\mathbf{k},d,\downarrow})$, where $l$ is the layer-index and $c_{\mathbf{k},s,\sigma,l}$ is electron annihilation operator with orbit $s$, spin $\sigma$ and layer $l$.
The interlayer coupling Hamiltonian $H_c(\mathbf{k})$ is given by 
\begin{equation}\label{eq:coupling_kdotp}
	H_{c}(\mathbf{k})=(B(\mathbf{k})+C(\mathbf{k})s_z+Dk_xs_y)\tau_x+(\gamma_0k_x\tau_y+\gamma_1k_x\tau_ys_z)+\alpha(\mathbf{k})s_x\tau_z+\beta(\mathbf{k})s_y\tau_y,
\end{equation}
where $B(\mathbf{k})=B_0+B_xk_x^2+B_yk_y^2$, $C(\mathbf{k})=C_0+C_xk_x^2+C_yk_y^2$, $\alpha(\mathbf{k})=\alpha_0+\alpha_xk_x^2+\alpha_yk_y^2$, $\beta(\mathbf{k})=\beta_0+\beta_xk_x^2+\beta_yk_y^2$. For simplicity,  only the spin-independent interlayer coupling is considered.

The value of model parameters, which are obtained from fitting the band structures of the bilayer 2M-WS2, are listed in the Supplementary Table \ref{table:2MWS2_parameter}. The chemcial potential $\mu$ is set to be 90 meV for the main text Fig.~\ref{fig:fig1}.

\begin{table}[h]
	\caption{List of model parameters (in units of meV) obtained from fitting the band structures of the bilayer 2M-WS2. }
	\begin{tabular}{cccccccc}
		\hline\hline
		$t_{0x}$&$t_{0y}$&$t_{2x}$&$t_{2y}$&$\delta$	&$m_{0x}$&$m_{0y}$&$m_{2x}$\\
		-3.21& 67.0991&-14.37&-6.19&-88.80&-117.99&-190.44&-0.70\\
		$m_{2y}$&$v_1$&$v_2$&$v_3$& $A_x$& $A_y$ & $A_z$&$B_0$\\
		-204.65&110.37&-18.37&-203.73&20.00&101.00&86.00&131.75\\
		$B_x$& $B_y$& $C_0$& $C_x$& $C_y$&$\alpha_0$&$\alpha_x$&$\alpha_y$\\
		
		-25.45&52.66&42.53&-0.34&22.88&-32.02&-3.87&99.16\\
		$\beta_0$&$\beta_x$&$\beta_y$&$D$&$\gamma_0$&$\gamma_1$&&\\
		-5.39&-11.53&77.81&-112.31&63.36&29.21&&\\
		\hline\hline
	\end{tabular}
    \label{table:2MWS2_parameter}
\end{table}

\section{Pairing correlations} 

Although we only consider the simplest conventional intra-orbital spin singlet pairing, i.e., $\Delta i\sigma_y$, here we show that the SOPC will induce some interorbital spin-triplet correlations, which also helps to enhance the upper critical field. To show this, 
we consider the following minimal model for a SOPC superconductor:
\begin{equation}
	H_N(\mathbf{k})=\epsilon_{\mathbf{k}}+\mathcal{M}_{\mathbf{k}}s_z+vk_xs_y+(A_xk_xs_x\sigma_y+A_yk_ys_x\sigma_x+A_zk_ys_x\sigma_z)+u_B \bm{B}\cdot\bm{\sigma},
\end{equation}
and the BdG Hamiltonian is
\begin{equation}
	H_{BdG}(\mathbf{k})=\begin{pmatrix}
		H_N(\mathbf{k},\bm{B})& \Delta i\sigma_y\\
		(\Delta i\sigma_y)^{\dagger}& -H_N^{*}(-\mathbf{k},\bm{B})
	\end{pmatrix}\label{noorbit}
\end{equation}

Let us identify the superconducting properties in terms of Green's function.
\begin{eqnarray}
	G_{\lambda\mu}(\bm{k},\tau)&&=-\braket{T_{\tau}\{c_{\bm{k},\lambda}(\tau)c^{\dagger}_{\bm{k},\mu}(0)\}},\\
	F_{\lambda\mu}(\bm{k},\tau)&&=\braket{T_{\tau}\{c_{\bm{k},\lambda}(\tau)c_{-\bm{k},\mu}(0)\}}.
\end{eqnarray}
We can rewrite the Green's function in the Matsubara frequency space: $G_{\lambda\mu}(\bm{k},i\omega_n)=\int_0^{\beta}d\tau e^{i\omega_n\tau}G_{\lambda\mu}(\bm{k},\tau)$ and $F_{\lambda\mu}(\bm{k},i\omega_n)=\int_0^{\beta}d\tau e^{i\omega_n\tau}F_{\lambda\mu}(\bm{k},\tau)$. The latter $F_{\lambda\mu}(\bm{k},i\omega_n)$ represents the pairing correlations we refer. These two Green's functions are related to the Gor'kov Green's function as
\begin{equation}
	\mathcal{G}(\bm{k},i
	\omega_n)=(i\omega_n-H_{BdG}(\bm{k}))^{-1}=\begin{pmatrix}
		G(\bm{k},i\omega_n)&-F(\bm{k},i\omega_n)\\-F^{\dagger}(\bm{k},i\omega_n)&-G^{T}(-\bm{k},-i\omega_n)
	\end{pmatrix}.
\end{equation}
Substitute the BdG Hamiltonian into Eq.~(\ref{noorbit}) and after some massage, we can parameterize the pairing correlation as 
\begin{equation}
	F(\bm{k},i\omega_n)=\Delta[\hat{C}_1(\bm{k},i\omega_n) + \hat{C}_2(\bm{k},i\omega_n)s_x\bm{d}(\bm{k})\cdot \bm{\sigma}] i\sigma_y\label{pairing_corr_supp}
\end{equation}
with the coefficients
\begin{align}
	\hat{C}_1(\bm{k},i\omega_n)&=\frac{1}{Q(\bm{k}, \omega_n)} \left[-D(\bm{k}, \omega_n) +2\epsilon_{\bm{k}} M_{\bm{k}}s_z +2\epsilon_{\bm{k}} vk_xs_y\right],\\
	\hat{C}_2(\bm{k},i\omega_n)&=\frac{2\epsilon_{\bm{k}}}{Q(\bm{k}, \omega_n)},\\
	Q(\bm{k}, \omega_n) &= 4\epsilon_{\bm{k}}^2 (v^2k_x^2+|\bm{Ak}|^2+M_{\bm{k}}^2)-D(\bm{k}, \omega_n)^2,\\
	D(\bm{k}, \omega_n) &= \omega_n^2+\epsilon_{\bm{k}}^2+M_{\bm{k}}^2+v^2k_x^2+|\bm{Ak}|^2 + \Delta^2,
\end{align}
where the SOPC vector $\bm{A}\bm{k}\equiv (A_yk_y, A_xk_x,A_zk_y)$.
Importantly, the triplet  vector is directly related to the SOPC: 
\begin{equation}
	\bm{d}(\bm{k})\approx (A_yk_y,A_xk_x,A_zk_y),
\end{equation}
where we have ignore the $O(\bm{k}^2)$ terms. Therefore, it can be seen that due to the presence of SOPC, the intraorbital spin-singlet pairing and the interorbital spin-triplet pairing are mixed with each other.
\section{The evaluation of pairing susceptibility in band basis}

In practice, it is more convenient to evaluate the pairing susceptibility in band basis. Specifically, we can write the pairing susceptibility $\chi^p_{\Gamma ,mm'}$ as 
\begin{eqnarray}
	\chi^p_{\Gamma ,mm'}&&=-\frac{1}{\beta}\sum_{n,\bm{p}}\text{Tr}[G_e(\bm{p},i\omega_n)(\Delta_{\Gamma,m}i\sigma_y)G_h(\bm{p},i\omega_n)(\Delta_{\Gamma,m'} i\sigma_y)^{\dagger}]\nonumber\\
	&&=\int \frac{d^2\bm{p}}{(2\pi)^2} \sum_{a,b}O^{\Gamma m}_{a,b}(\bm{p})O^{\Gamma m'\,\dagger}_{a,b}(\bm{p})\frac{1-f(E_a(\bm{p}))-f(E_b(-\bm{p}))}{E_a(\bm{p})+E_b(-\bm{p})}.
\end{eqnarray}
Here, the single particle electron Green's function $G_e(\bm{p},i\omega_n)=(i\omega_n-H_0(\bm{p}))^{-1}$ and hole Green's function $G_h(\bm{p},i\omega_n)=(i\omega_n+H^*_0(-\bm{p}))^{-1}$, the overlap function $O^{\Gamma m}_{a,b}(\bm{p})=\braket{u_{a,\bm{p}}|\Delta_{\Gamma,m}i\sigma_y|\nu_{b,-\bm{p}}}$ with $\ket{u_{a,\bm{p}}}$,$\ket{\nu_{b,\bm{p}}}$ being eigenvectors of $H_0(\bm{p})$ satisfying $H_0(\bm{p})\ket{u_{a,\bm{p}}}=E_a(\bm{p})\ket{u_{a,\bm{p}}}, H_0^*(\bm{p})\ket{\nu_{b,\bm{p}}}=E_b(\bm{p})\ket{\nu_{b,\bm{p}}}$, $a,b$ are the band indices.

\section{two-fold Andreev reflection amplitude in FM/SOPC SC junction}

\begin{figure}
	\centering
	\includegraphics[width=1\linewidth]{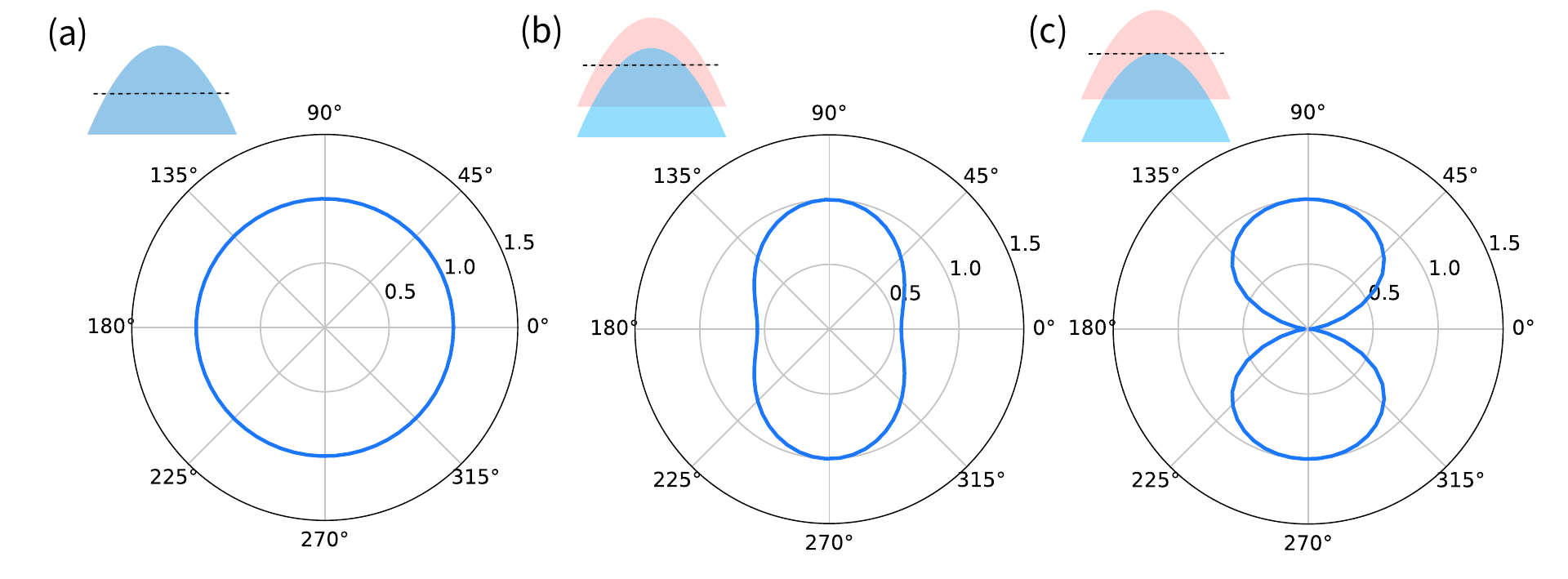}
	\caption{(a)-(c) The angular dependence of in-gap normalized tunnel conductance $G$, where the angle represents the in-plane magnetization direction, in the FM/1T'-WTe$_2$ SC tunneling junction. We set $|\bm{m}|/t_{1p}$ = 0, 0.6, 0.8 in (a)-(c), respectively. The major spin and minor spin population are illustrated in each case. Again, it can be clearly seen that the Andreev reflection amplitude exhibits a two-fold feature when the magnetization amplitude $\bm{m}$ is finite in the ferromagnetic metal lead. The two-fold anisotropic feature becomes significantly more salient when ferromagentic metal lead is close to a half-metal in (c).}
	\label{fig:S1}
\end{figure}

In this section, we present details of the tight-binding model used to study the tunneling spectrum in Fig.~\ref{fig:fig4} in the main text.
In the Nambu basis $(c_{\bm{k},p,\uparrow},c_{\bm{k},p,\downarrow}, c_{\bm{k},d,\uparrow},c_{\bm{k},d,\downarrow}, c^{\dagger}_{-\bm{k},p,\uparrow}, c^{\dagger}_{-\bm{k},p,\downarrow}, c^{\dagger}_{-\bm{k},d,\uparrow}, c^{\dagger}_{-\bm{k},d,\downarrow})^{T}$, 
where $c^{\dagger}_{\bm{k},l,\sigma}$ ($l=p,d,\sigma = \uparrow, \downarrow$) creates a Bloch state formed by linear combinations of Wannier orbital of character $l$ and spin $\sigma$, the momentum-space tight-binding Hamiltonian $\hat{H}^{TB}_{BdG}(\bm{k})$ for the SOPC superconductor under the superconducting $\hat{\Delta}$ reads:
\begin{eqnarray}
	\hat{H}^{TB}_{BdG}(\bm{k}) &=& \sum_{\bm{k}, mn} c^{\dagger}_{\bm{k},m} H^{TB}_{0, mn} (\bm{k}) c_{\bm{k},n} + \Delta (c^{\dagger}_{\bm{k},p,\uparrow} c^{\dagger}_{-\bm{k},p,\downarrow} + c^{\dagger}_{\bm{k},d,\downarrow} c^{\dagger}_{-\bm{k},d,\uparrow} + h.c.).
\end{eqnarray}
Here, $m,n = (l, \sigma)$ label the index for different Wannier orbitals with $l= p,d$, $\sigma=\uparrow, \downarrow$. $H^{TB}_{0} (\bm{k})$ is an $4\times4$ matrix given by:
\begin{eqnarray}\label{eq:TB_monolayer}
	H^{TB}_{0} (\bm{k}) &=& \begin{pmatrix}
		E_p (\bm{k}) - \mu & 0 & -i v\sin(k_x a) + A_z\sin(k_y b) & -i A_x\sin(k_x a) + A_y\sin(k_y b)\\
		& E_p (\bm{k}) - \mu & i A_x\sin(k_x a) + A_y\sin(k_y b) & -i v\sin(k_x a) - A_z\sin(k_y b)\\
		&   & E_d(\bm{k})  - \mu   & 0\\
		h.c.&  &  &     E_d(\bm{k}) - \mu
	\end{pmatrix},
\end{eqnarray}

\begin{align}
	E_p (\bm{k}) =\,& 2\left\{  t_{1p} [\cos(k_x a) - 1] + t_{2p}[\cos(k_y b) - 1] \right\} - \mu_p, \nonumber \\
	E_d (\bm{k}) =\,& 2\left\{ t_{1d} [\cos(k_x a) - 1] + t_{2d}[\cos(k_y b) - 1] + t'_{2d}[\cos(2 k_y b) - 1]\right\} - \mu_d.
\end{align}
Here, a monolayer Hamiltonian of 2M-WS$_2$ is used for simplicity. The parameters above are tabulated in Table \ref{table:TB_2MWS2_parameter}. 
It can be verified that $H^{TB}_{0} (\bm{k})$ reduces to the $\bm{k} \cdot \bm{p}$ model near the $\Gamma$-point in Eq. (\ref{eq:monolayer_kdotp}) in the continuum limit $a,b \rightarrow 0$.

The ferromagentic lead Hamiltonian can be written as
\begin{equation}
	H_{\text{lead}}=\sum_{k_x,k_y,s,s'} \psi^{\dagger}_{k_x,k_y,s}\{[ 2t_L\cos(k_xa)+2t_L \cos(k_yb)-\mu_L]\sigma_{ss'}^0 + \bm{m}\cdot\bm{\sigma}_{ss'}\}\psi_{k_x,k_y,s'}
\end{equation}
We can the Fourier transform: $\psi_{m,n,s} = \frac{1}{\sqrt{N}}\sum_{k_x,k_y} e^{-i m k_xa-ink_yb } \psi_{k_x,k_y,s},c_{m,n,p(d),s} = \frac{1}{\sqrt{N}}\sum_{k_x,k_y} e^{-i m k_xa-ink_yb } c_{k_x,k_y,p(d),s} $ and change above Hamiltonian as the lattice model. The coupling Hamiltonian between the lead and the SOPC superconductor is written as
\begin{equation}
	H_{\text{couple}}=\sum_{m,n,s}-(t_c\psi^{\dagger}_{m,n,s}c_{m,n,p, s}+t_c\psi^{\dagger}_{m,n,s}c_{m,n,d, s})+h.c.
\end{equation}

The tunneling spectrum is calculated by the recursive Green's function (c.f. Ref.~\cite{Manna2020_supp}):
 \begin{equation}
 	G(E)=\frac{e^2}{h}\text{Tr}[I-r^{\dagger}_{ee}(E)r_{ee}(E)+r^{\dagger}_{he}(E)r_{he}(E)]. \end{equation}
Here $r_{ee}/r_{he}$ are the normal reflection/Andreev reflection amplitude respectively. In main text Fig. ~\ref{fig:fig4}, we set $\mu = 90$ meV, $\Delta_1=6$ meV, $t_L=4|\bm{m}|$, $t_c=t_{1p}/3$, and the magnetization vector $\bm{m}=|\bm{m}|(\cos\theta,\sin\theta,0)$, and $E=0$. The magnetization amplitude $|\bm{m}|$ for main text Fig. ~\ref{fig:fig4}(b)-(d) are $|\bm{m}|=0,\,0.75,\,0.8\,|t_{1p}|$, respectively. 

To show that spin-triplet pairing generally exists in SOPC superconductor, we adopt 1T'-WTe$_2$ parameters from Ref.~\cite{Law2020_supp} (tabulated in Table.~\ref{table:TB_1TWTe2_parameter}) and perform the same caluculation. In Fig.~\ref{fig:S1}, we set $\mu = 60$ meV. The other parameters of the lead and the coupling Hamiltonian take the same form as before. 

\begin{table}[h]
	\caption{2M-WS$_2$ tight-binding parameters in $H^{TB}_{0} (\bm{k})$ in Eq.~(\ref{eq:TB_monolayer}) (in units of meV). Lattice constants: $a$ = 5.71 \AA, $b$ = 3.23 \AA.}
    \label{table:TB_2MWS2_parameter}
	\begin{tabular}{ccccccccccc}
 		\hline\hline
 		$\mu_p$ & $\mu_d$ & $t_{1p}$ & $t_{2p}$ & $t_{1d}$ & $t_{2d}$ &$t'_{2d}$ & $v$ &$A_x$ & $A_y$ & $A_z$ \\
 		\hline
 		-88.8 & 88.8 & -121.2 & -123.34 & -114.78 & 337.84 & 148.85 & 110.37 & 20 &101 &86 \\
 		\hline\hline
 	\end{tabular}	
\end{table}

\begin{table}[h]	
    \caption{1T$'$-WTe$_2$ tight-binding parameters in $H^{TB}_{0} (\bm{k})$ in Eq.~(\ref{eq:TB_monolayer}) (in units of meV). Lattice constants: $a=6.31${\AA}, $b=3.49${\AA}.}
    \label{table:TB_1TWTe2_parameter}
 	\begin{tabular}{ccccccccccc}
 		\hline\hline
 		$\mu_p$ & $\mu_d$ & $t_{1p}$ & $t_{2p}$ & $t_{1d}$ & $t_{2d}$ &$t'_{2d}$ & $v$ &$A_x$ & $A_y$ & $A_z$ \\
 		\hline
 		-1390 & 62 & 626 & 1517 & -60 & -387 & 150 & 371 & 27 &163 &20\\
 		\hline\hline
 	\end{tabular}
\end{table}

\end{document}